\documentclass{PHYEAUTH}
\usepackage{graphicx}
\usepackage{amsmath}
\usepackage{amssymb}

\begin{document}

\begin{frontmatter}

\title{
Thermopower of Kondo Effect in Single Quantum Dot Systems with Orbital at Finite Temperatures
}

\author[address1]{R. Sakano\thanksref{thank1}},
\author[address1]{T. Kita}
and
\author[address1,address2]{N. Kawakami}

\address[address1]{Department of Applied Physics, Osaka University, Suita, Osaka 565-0871, Japan}

\address[address2]{Department of Physics, Kyoto University, Kyoto 606-8502, Japan}

\thanks[thank1]{
E-mail: sakano@tp.ap.eng.osaka-u.ac.jp}

\begin{abstract}
We investigate the thermopower due to the orbital Kondo effect in a single quantum dot system by means of the noncrossing approximation.
It is elucidated  how the asymmetry of tunneling resonance due to 
the orbital Kondo effect affects the thermopower under gate-voltage 
and magnetic-field control.

\end{abstract}

\end{frontmatter}

\section{Introduction}%
The Kondo effect due to magnetic impurity scattering in metals is a well known and widely studied phenomenon \cite{book:hewson}.
The effect has recently received much renewed  attention since it was found that the Kondo effect significantly influences the conductance in quantum dot (QD) systems
\cite{pap:D.GG}.
A lot of tunable parameters in QD systems have made it possible to systematically investigate electron correlations.
In particular, high symmetry in shape of QDs gives rise to the orbital properties, which has stimulated extensive studies on the conductance due to the orbital Kondo effect
\cite{pap:sasaki_st,pap:st_Eto,pap:Sasaki2,pap:pjh,pap:Choi,pap:sakano}.
The thermopower we study in this paper is another important transport quantity, which gives complementary information on the density of states to the conductance measurement: the thermopower can sensitively probe the asymmetric nature of the tunneling resonance around the Fermi level.
So far, a few theoretical studies have been done on the thermopower in QD systems
\cite{pap:Beenakker,pap:boese,pap:Turek,pap:tskim,pap:Matveev,pap:BDong,pap:Krawiec,pap:Donabidowicz}. A recent observation of
 the thermopower due to the spin Kondo effect in a lateral QD system
\cite{pap:Scheibner} 
 naturally motivates us to theoretically explore this transport 
quantity in more detail.
Here, we discuss how the asymmetry of tunneling resonance due to the orbital Kondo effect affects the thermopower under gate-voltage and magnetic-field control. By employing the noncrossing approximation (NCA) for the  Anderson model with finite Coulomb repulsion, we especially investigate the Kondo effect of QD for several electron-charge regions.

\section{Model and Calculation}%
\begin{figure}[bt]
\begin{center}\leavevmode
\includegraphics[width=0.5\linewidth]{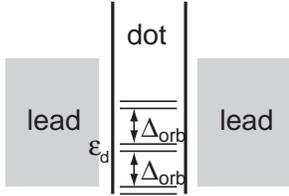}
\caption{
Energy-level scheme of a single QD system with three orbitals coupled to two leads.
}
\label{fug:sch}
\end{center}
\end{figure}

Let us consider a single QD system with $N$-degenerate orbitals in equilibrium, as shown in Fig. \ref{fug:sch}.
The energy levels of the QD are assumed to be
\begin{eqnarray}
&&\varepsilon_{\sigma l} = \varepsilon_d + l \Delta_{orb}, \\
&&l=-(N_{orb}-1)/2,-(N_{orb}-3)/2, \cdots, (N_{orb}-1)/2 \nonumber
\end{eqnarray}
where $\varepsilon_d$ denotes the center of the energy levels and $\sigma$ ($l$) represents spin (orbital) index and $N_{orb}$ represents the degree of the orbital degeneracy.
The energy-level splitting between the orbitals $\Delta_{orb}$ is induced in the presence of magnetic field $B$; $\Delta_{orb} \propto B$.
In addition, the Zeeman splitting is assumed to be much smaller than the orbital splitting, so that we can ignore the Zeeman effect.
In practice, this type of orbital splitting has been experimentally realized as Fock-Darwin states in vertical QD systems or clockwise and counterclockwise states in carbon nanotube QD systems.
Our QD system is described by the multiorbital Anderson impurity model,
\begin{eqnarray}
{\cal H} &=& {\cal H}_l + {\cal H}_d + {\cal H}_{t} \label{eq:hamiltonian} \\
{\cal H}_l &=& \sum_{k \sigma l} \varepsilon_{k \sigma l} c^{\dagger}_{k \sigma l} c_{k \sigma l}, \\
{\cal H}_d &=& \sum_{k \sigma l} \varepsilon_{\sigma l} d^{\dagger}_{\sigma l} d_{\sigma l}
+ U \sum_{\sigma l \neq \sigma' l'} n_{\sigma l} n_{\sigma' l'} \nonumber \\
&& \qquad -J \sum_{l \neq l'} \mbox{\boldmath$S$}_{dl} \cdot \mbox{\boldmath$S$}_{dl'} , \\
{\cal H}_{t} &=& V \sum_{k \sigma } \left( c^{\dagger}_{k \sigma l} d_{\sigma l} + \mbox{H. c.} \right),
\end{eqnarray}
where $U$ is the Coulomb repulsion and $J(>0)$ represents the Hund coupling in the QD.

The non-equilibrium Green's function technique allows us to study 
general transport properties, which gives the expression for the T-linear
thermopower as \cite{pap:BDong},
\begin{eqnarray}
S=-(1/eT) ({\cal L}_{12}/{\cal L}_{11}),
\end{eqnarray}
with the linear response coefficients,
\begin{eqnarray}
{\cal L}_{11} &=& \frac{\pi T}{h} \Gamma \sum_{\sigma l} \int d\varepsilon \, \rho_{\sigma l}(\varepsilon) \left( - \frac{\partial f(\varepsilon)}{\partial \varepsilon} \right), \\
{\cal L}_{12} &=& \frac{\pi T}{h} \Gamma \sum_{\sigma l} \int d\varepsilon \, \varepsilon \rho_{\sigma l} (\varepsilon) \left( - \frac{\partial f(\varepsilon)}{\partial \varepsilon} \right),
\end{eqnarray}
where $\rho_{\sigma l}(\varepsilon)$ is the density of states for the electrons with spin $\sigma$ and orbital $l$ in the QD and $f(\varepsilon)$ is the Fermi distribution function. In order to obtain the thermopower it is necessary to evaluate $\rho_{\sigma l}(\varepsilon)$.

We exploit the NCA method to treat the Hamiltonian (\ref{eq:hamiltonian})
\cite{pap:Bickers,pap:Pruschke}.
The NCA is a self-consistent perturbation theory, which summarizes a specific series of expansions in the hybridization $V$.
This method is known to give physically sensible results at temperatures around or higher than the Kondo temperature.
The NCA basic equations can be obtained in terms of coupled equations for the self-energies $\Sigma_m(z)$ of the resolvents $R_m(z)=1/[z-\varepsilon_m - \Sigma_m(z)]$,
\begin{eqnarray}
\Sigma_m(z) &=& \frac{\Gamma}{\pi} \sum_{m'} \sum_{\sigma l} \left[ \left( M^{\sigma l}_{m' m} \right)^2 + \left( M^{\sigma l}_{m m'} \right)^2 \right] \nonumber \\
&& \qquad \times \int d\varepsilon R_{m'}(z+\varepsilon)f(\varepsilon),
\end{eqnarray}
where the index $m$ specifies the eigenstates of ${\cal H}_d$ 
and the mixing width is $\Gamma=\pi \rho_c V^2$.
The coefficients $M_{mm'}^{\sigma l}$ are determined by the expansion coefficients of the Fermion operator $d_{\sigma l}^{\dagger}=\sum_{mm'} M_{mm'}^{\sigma l} | m \rangle \langle m' |$.
We compute the density of states by this method to investigate the thermopower.

\section{Results}

\subsection{Gate voltage control}%
\begin{figure}[bt]
\begin{center}\leavevmode
\includegraphics[width=6cm]{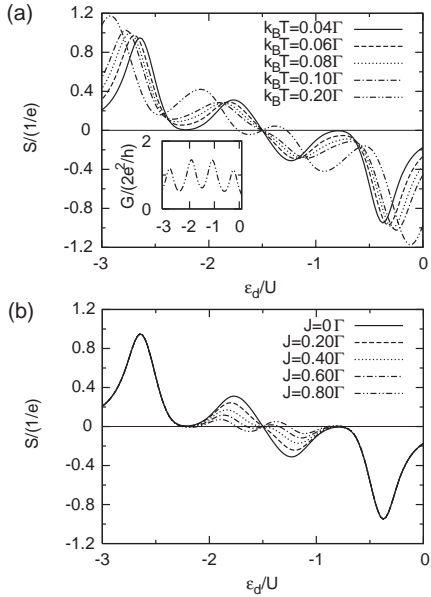}
\caption{
The thermopower for the two orbital QD system with finite Coulomb repulsion $U=8\Gamma$ as a function of the energy level of the QD.
(a) The temperature dependence for $J=0$. The inset shows the conductance as a function of the dot level at $k_BT=0.20\Gamma$ (Coulomb resonance peaks).
(b) The Hund-coupling dependence for $k_BT=0.04\Gamma$.
}
\label{fig:double_vS}
\end{center}
\end{figure}

The thermopower for two orbitals is shown in Fig. \ref{fig:double_vS} 
as a function of the energy level $\varepsilon_d$ (gate-voltage control).
There are four Coulomb peaks around $-\varepsilon_d/U \sim 0,1,2,3$ at high temperatures (see the inset of Fig. \ref{fig:double_vS}(a)).
As the temperature decreases, the thermopower in the region of $-1<\varepsilon_d/U < 0 (-3<\varepsilon_d/U < -2)$ with $n_d \sim 1 (3)$ is dominated  by the 
{\it SU}(4) Kondo effect.
The thermopower has negative values in the region $-1<\varepsilon_d/U < 0$,
implying that the effective tunneling resonance, such as the 
Kondo resonance, is located above the Fermi level.
At low enough temperatures, the {\it SU}(4) Kondo effect is enhanced with decrease of energy level down to $\varepsilon_d/U=-1/2$, which results in the enhancement of the thermopower.
However, if the temperature of the system is larger than the {\it SU}(4) Kondo temperature, the Kondo effect is suppressed and the thermopower has 
a minimum in the  regime $-1/2 <\varepsilon_d/U <0$.
As the energy level further decreases, the {\it SU}(4)
Kondo effect and the resulting thermopower are both suppressed.
Note that the Hund coupling hardly affects the thermopower because of $n_d \sim 1$ in this regime, as shown in Fig. \ref{fig:double_vS} (b).
Since the region of $-3<\varepsilon_d/U < -2$ can be related to
 $-1<\varepsilon_d/U < 0$ via an electron-hole transformation,
we can directly apply the above discussions on the {\it SU}(4) Kondo effect 
to the former region by changing the sign of the thermopower.

\begin{figure}[bt]
\begin{center}\leavevmode
\includegraphics[width=5.5cm]{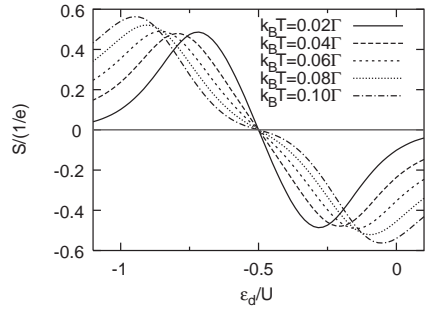}
\caption{The thermopower due to the ordinary spin Kondo effect as a function of the dot level. We set $U=6\Gamma$.}
\label{fig:single_vS}
\end{center}
\end{figure}

Let us now turn to the region of $-2<\varepsilon_d/U<-1$, where $n_d \sim 2$.
At $J=0$, the Kondo effect due to six-fold degenerate states occurs.
Although the resulting Kondo effect is strongly enhanced around
$\varepsilon_d/U=-3/2$ in this case,
  the thermopower is almost zero because the Kondo resonance
 is located just at the Fermi level.
Therefore, when the dot level is changed,
the position of the Kondo resonance
is shifted  across the Fermi level, which causes the
sign change of the thermopower.
Around $\varepsilon_d/U=-3/2$, even small perturbations could easily change the sign of the thermopower at low temperatures. Note that
these properties are quite similar to those for the ordinary spin Kondo
effect shown in Fig. \ref{fig:single_vS}, because
 the filling is near half in both cases.
For large Hund couplings $J$, the triplet Kondo effect 
is realized and the resulting Kondo temperature is very small,
so that the thermopower shown in Fig. \ref{fig:double_vS}(b)
is dramatically suppressed.

\subsection{Magnetic field control}%
\begin{figure}[bt]
\begin{center}\leavevmode
\includegraphics[width=6cm]{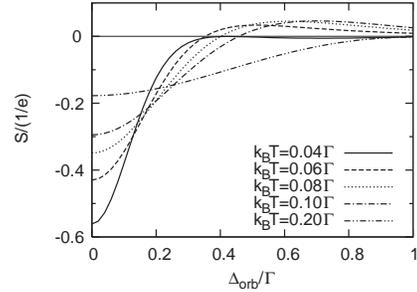}
\caption{
The thermopower for the two orbital QD system, in case of
$\varepsilon_d=-U/2$, as a function of orbital splitting $\Delta_{orb}$.
We set $U=8\Gamma$.
 }
\label{fig:double_kS}
\end{center}
\end{figure}

Let us now analyze the effects of orbital-splitting caused by 
 magnetic fields. The computed 
thermopower for  $\varepsilon_d/U=-1/2$ is shown 
in Fig. \ref{fig:double_kS} as a function of 
the orbital splitting $\Delta_{orb}$.
It is seen that magnetic fields dramatically suppress the thermopower,
which is caused by the following mechanism.
In the presence of magnetic fields, the Kondo effect
changes from the {\it SU}(4) orbital type to the {\it SU}(2) spin
type because the orbital degeneracy is lifted.
As a consequence, the resonance peak approaches the Fermi level and
the effective Kondo temperature is reduced, so that
the thermopower at finite temperatures is reduced
in the presence of magnetic fields.

Note that, in our model, magnetic fields change the lowest energy level 
$\varepsilon_{\sigma -\frac{1}{2}}$ from $-U/2$ to $-(U+\Delta_{orb})/2$.
Accordingly, the peak position of the renormalized resonance shifts 
downward across the Fermi level (down to a little 
below the Fermi level).
Thus, the large negative thermopower changes to  a small positive one 
as the magnetic field increases at low temperatures.
In strong fields, the effective Kondo resonance is located
around the Fermi level with symmetric shape, so that
 even small perturbations could give rise to a large value 
of thermopower with either negative or positive sign.

Finally a brief comment is in order for other choices of the parameters.
The thermopower for $\varepsilon_d/U=-5/2$ shows similar 
magnetic-field dependence to the $\varepsilon_d/U=-1/2$ case except that
 its sign is changed.
For $\varepsilon_d/U=-3/2$, the thermopower is almost zero and independent
 of magnetic fields, because the Kondo resonance is pinned at the Fermi 
level and gradually disappears with increase of magnetic fields.

\section{Summary}%
We have studied the thermopower for the two-orbital QD system 
under gate-voltage and magnetic-field control.
In particular, making use of the NCA method for the Anderson model 
with finite Coulomb repulsion, we have systematically investigated 
the low-temperature properties for several electron-charge regions.
It has been elucidated  how the asymmetric nature of the resonance due
to the orbital Kondo effect controls the magnitude and the sign of the
thermopower at low temperatures.

For $\varepsilon_d/U\sim-1/2$ ($\varepsilon_d/U\sim -5/2$), where $n_d \sim1 (3)$, the {\it SU}(4) Kondo effect is dominant and the corresponding thermopower is enhanced.
These two regions are related to each other via an electron-hole transformation, which gives rise to an opposite sign of the thermopower.
In addition, magnetic fields change the Kondo effect to an {\it SU}(2) type, 
resulting in two major effects: the effective resonance 
position approaches the Fermi level and  the Kondo temperature is decreased. Therefore, the reduction 
of the thermopower  occurs in the presence of magnetic fields.

For $\varepsilon_d/U\sim-3/2$, where $n_d \sim 2$, the Kondo effect due to six-fold degenerate states occurs for $J=0$. However,
the thermopower is strongly reduced because the resonance peak is located
near the Fermi level.
When the Hund coupling is large, the triplet Kondo effect is dominant.
The resulting small Kondo temperature suppresses the thermopower 
around $\varepsilon_d/U \sim -3/2$ at finite temperatures.
In this region, magnetic fields do not affect the asymmetry of the 
resonance peak and the resulting thermopower remains almost zero 
because the filling is fixed.

\section*{Acknowledgement}%
We thank S. Tarucha, A. C. Hewson, A. Oguri and S. Amaha for
 valuable discussions.
RS was supported by the Japan Society for the Promotion of Science.



\begin{thebibliography}{00}
\bibitem{book:hewson}
A. C. Hewson, {\it The Kondo Problem to Heavy Fermions}
(Cambridge University Press, Cambridge, 1997).

\bibitem{pap:D.GG}
D. Goldhaber-Gordon, {\it et al.},
Nature, {\bf 391} (1998) 156.

\bibitem{pap:sasaki_st}
S. Sasaki, {\it et al.},
Nature, {\bf 405} (2000) 764.

\bibitem{pap:st_Eto}
M. Eto, {\it et al.},
Phys. Rev. Lett. {\bf 85} (2000) 1306.

\bibitem{pap:Sasaki2}
S. Sasaki, {\it et al.},
Phys. Rev. Lett. \textbf{93} (2004) 17205.

\bibitem{pap:pjh}
P. Jarillo-Herrero, {\it et al.},
Nature, {\bf 434} (2005) 484.

\bibitem{pap:Choi}
M.-S. Choi, {\it et al.},
Phys. Rev. Lett. \textbf{95} (2005) 067204.

\bibitem{pap:sakano}
R. Sakano, {\it et al.},
Phys. Rev. B {\bf 73} (2006) 155332.

\bibitem{pap:Beenakker}
C. W. J. Beenakker,
Phys. Rev. B {\bf 46} (1992) 9667.

\bibitem{pap:boese}
D. Boese, {\it et al.},
Euro. Phys. Lett. {\bf 56} (2001) 576.

\bibitem{pap:Turek}
M. Turek, {\it et al.},
Phys. Rev. B {\bf 65} (2002) 115332.

\bibitem{pap:tskim}
T.-S. Kim, {\it et al.},
Phys. Rev. Lett. {\bf 88} (2002) 136601.

\bibitem{pap:Matveev}
K. A. Matveev, {\it et al.},
Phys. Rev. B {\bf 66} (2002) 45301.

\bibitem{pap:BDong}
B. Dong, {\it et al.},
J. Phys. C {\bf 14} (2002) 11747.

\bibitem{pap:Krawiec}
M. Krawiec, {\it et al.},
Phys. Rev. B {\bf 73} (2006) 75307.

\bibitem{pap:Donabidowicz}
A. Donabidowicz, {\it et al.},
preprint, cond-mat/0701217, (2007).

\bibitem{pap:Scheibner}
R. Scheibner, {\it et al.},
Phys. Rev. Lett. {\bf 95} (2005) 176602.

\bibitem{pap:Bickers}
N. E. Bickers,
Rev. Mod. Phys. \textbf{59}, (1987) 845.

\bibitem{pap:Pruschke}
Th. Pruschke, {\it et al.},
Z. Phys. {\bf 74} (1989) 439.

\bibitem{pap:}
W. Izumida, {\it et al.},
Phys. Rev. Lett. {\bf 87} (2001) 216803.

\end{thebibliography}
\end{document}